\documentclass[12pt]{article}

\usepackage{amssymb,amsmath,amsfonts,eurosym,geometry,ulem,graphicx,caption,color,setspace,sectsty,comment,footmisc,caption,pdflscape,subcaption,array,hyperref,booktabs, threeparttable}

\usepackage[affil-it]{authblk}
\usepackage[
backend=biber,
style=numeric-comp,
maxbibnames=6,
maxcitenames=2,
terseinits=true,
firstinits=true,
doi=true,
isbn=false,
url=false,
eprint=false,
maxbibnames=99,
date=year,
sorting=none
]{biblatex}

\newcommand\Tstrut{\rule{0pt}{2.6ex}}         
\newcommand\Bstrut{\rule[-0.9ex]{0pt}{0pt}}   

\newcolumntype{L}[1]{>{\raggedright\let\newline\\arraybackslash\hspace{0pt}}m{#1}}
\newcolumntype{C}[1]{>{\centering\let\newline\\arraybackslash\hspace{0pt}}m{#1}}
\newcolumntype{R}[1]{>{\raggedleft\let\newline\\arraybackslash\hspace{0pt}}m{#1}}

\AtEveryBibitem{%
  \clearfield{note}%
  \clearlist{language}
}

\DeclareNameAlias{default}{last-first}

\newbibmacro*{in:}{}

\begin{document}

\renewcommand\Affilfont{\small \itshape}

\begin{titlepage}
\title{Outcome coding choice in randomized trials of programs to reduce violence.\thanks{Funding for this study was provided by a SVRI/World Bank Group Development Marketplace Award for Innovation on GBV prevention. We thank the participants at the SVRI Forum 2022 for helpful comments.}}
\author[1]{Christopher Boyer\thanks{email: \href{mailto:cboyer@g.harvard.edu}{cboyer@g.harvard.edu}}}
\author[2]{Sangeeta Chatterji}
\author[3]{Jasper Cooper}
\author[4]{Lori Heise}
\affil[1]{Department of Epidemiology, Harvard School of Public Health, Boston, MA, USA.}
\affil[2]{School of Social and Political Science, University of Edinburgh, Edinburgh, UK}
\affil[3]{Department of Political Science, University of California, San Diego, La Jolla, CA, USA}
\affil[4]{Department of Population, Family and Reproductive Health, Johns Hopkins Bloomberg School of Public Health, Baltimore, MD, USA.}
\date{\today}
\maketitle

\begin{abstract}
\noindent Over the last decade, the number of randomized trials of programs to reduce intimate partner violence (IPV) has grown precipitously. However, most trials continue to measure and code violence using standards originally designed for global prevalence surveys. This choice may have consequences in terms of bias, power, and efficiency of trial estimates and may limit what we can learn about how programs are working. In this paper, we return to first principles to develop a generative model for violence reduction. We then use this model to better understand trade-offs in outcome coding choices via simulation. We re-analyze results from seven recent trials in Southern and Eastern Africa to highlight some of our findings. We conclude with a discussion of key take-aways for trialists. \\
\vspace{0in}\\
\noindent\textbf{Keywords:} measurement, intimate partner violence, simulation, outcome coding, randomized controlled trial \\

\bigskip
\end{abstract}
\setcounter{page}{0}
\thispagestyle{empty}
\end{titlepage}
\pagebreak \newpage

\doublespacing

\section{Introduction} \label{sec:introduction}
Globally, nearly a third of women report experiencing violence perpetrated by an intimate partner \cite{devries_global_2013} with far-reaching consequences for their health, their families, their opportunities, and their livelihoods \cite{ellsberg_intimate_2008}. This has prompted interest among funders, practitioners, and policy-makers to develop programs to reduce or prevent violence and in a research agenda to better understand empirically what types of programs work in reducing violence \cite{heise_what_2011,jewkes_elements_2021}. Consequently, over the last two decades, there has been a surge in randomized evaluations of violence reduction programs, particularly in lower and middle income countries \cite{bourey_systematic_2015, abramsky_findings_2014, hidrobo_effect_2016, jewkes_impact_2008, pronyk_effect_2006, wagman_effectiveness_2015,boyer_religious_2022,doyle_gender-transformative_2018,dunkle_effective_2020, sharma_effectiveness_2020, kapiga_social_2019}. Most of these evaluations measure violence using a standardized instrument, based on the Conflict Tactics Scale (CTS)  \cite{straus_measuring_1979, straus_revised_1996}, as well as a standardized outcome coding scheme originally developed for large cross-sectional prevalence assessments. While this standardization allows for comparability across studies, to date, little work has been done to understand whether this coding choice is optimal in the context of a randomized evaluation.  

In this paper, we explore the consequences of this coding choice with respect to statistical bias and efficiency and we compare it to several alternatives, first via simulation, and then by re-analyzing data from several recent trials. To do so we return to the original conception of the CTS in the sociology literature to develop a generative model of violence. We also use potential outcomes and the Neyman-Rubin causal model \cite{splawa-neyman_application_1990, rubin_estimating_1974} to formalize reductions in violence in the context of randomized evaluations and to think structurally about how intervention effects operate. We then use this theory to simulate data from hypothetical trials under different violence reduction regimes and then compare outcome coding strategies. Finally, we re-analyze data from seven recent trials in Southern and Eastern Africa using alternative coding strategies and re-interpret them in light of the insights gleaned from simulations. 

\subsection{The Conflict Tactics Scale}\label{sec:cts}
Studies that measure intimate partner violence generally rely on some version of the Conflict Tactics Scale (CTS) to quantify a participant's experience of violence with perhaps the most widely used version being the standardized questionnaire developed by the World Health Organization \cite{world_health_organization_who_2005}. The origins of the instrument are rooted in conflict theory which posits that conflict is an inevitable part of human relationships, but that the tactics employed to deal with conflict vary. Among these tactics are included those that involve physical force, coercion, or verbal aggression, which we may define as ``violent''. However, the instrument consciously disassociates these tactics from their personal or social meaning as ``violence'' by asking respondents to report the frequency that they experienced specific acts rather than how often they experience ``violence''. This allows for comparable objective assessment of tactics used during conflict even when definitions of violence may vary from person to person.

The original CTS instrument was designed to assess both perpetration as well as experiences of violence and to capture family violence more broadly including violence directed against children. However, in evaluations of intimate partner violence reduction programs an abbreviated version is often used as the focus is typically limited to violence experienced by women, although sometimes supplemented with male reports of perpetration. The violence items in the scale were chosen to capture different latent constructs. In the original scale, these items loaded on two latent factors representing psychological aggression and physical assault. A revised scale in 1996 added items representing sexual coercion and also introduced a shortened version which was the basis for the WHO questionnaire. In more recent literature the three latent factors into which items are grouped are more commonly referred to as emotional violence, physical violence, and sexual violence. Table \ref{tab:cts} below gives examples of the items in the scale from the WHO questionnaire. Items can sometimes be added or deleted or adapted to local context, however the basic structure is largely the same. All items in the scale refer to the same defined recall period, usually 12 months, although this can vary in randomized evaluations where repeated assessments may be made and where interest is generally in violence experienced since the start of a violence reduction program.

\begin{table}[t]
\centering
\footnotesize{
    \caption{Example of CTS style questions for measuring violence adapted from the WHO Domestic Violence questionnaire.}
    \label{tab:cts} 
    \begin{tabular}[t]{p{0.5cm}p{10cm}p{0.5cm}p{0.5cm}p{0.5cm}p{0.5cm}}
    \hline \hline 
    \multicolumn{6}{p{14cm}}{\hspace{2em} No matter how well a couple gets along, there are times when they disagree on major decisions, get annoyed about something the other person does, or just have spats or fights because they're in a bad mood or tired or for some other reason. They also use many different ways of trying to settle their differences. I'm going to read a list of some things that you and your (husband/partner) might have done when you had a dispute, and would first like you to tell me how often your (husband/partner) has done them in the past.} \vspace{1em} \Tstrut\Bstrut \\
    \multicolumn{2}{l}{In the past 12 months, how often has your partner...} & \rotatebox{90}{Never} & \rotatebox{90}{Once} & \rotatebox{90}{A few times (2-4)} & \rotatebox{90}{Many times (5+)} \Tstrut\Bstrut\\
    \hline
    1. &  insulted you or made you feel bad about yourself & 0 & 1 & 2 & 3 \Tstrut\\
    2. &  belittled or humiliated you in front of others? & 0 & 1 & 2 & 3\\
    3. &  did things to scare or intimidate you on purpose? & 0 & 1 & 2 & 3\\
    4. &  threatened to hurt you or someone you care about? & 0 & 1 & 2 & 3\\
    5. &  slapped you or thrown something at you that could hurt you? & 0 & 1 & 2 & 3 \Tstrut\\
    6. &  pushed you or shoved you or pulled your hair? & 0 & 1 & 2 & 3\\
    7. &  hit you with his fist or with something else that could hurt you? & 0 & 1 & 2 & 3\\
    8. &  kicked you, dragged you or beaten you up? & 0 & 1 & 2 & 3\\
    9. &  choked or burnt you on purpose? & 0 & 1 & 2 & 3\\
    10. &  threatened you with or actually used a gun, knife or other weapon against you? & 0 & 1 & 2 & 3\\
    11. &  physically forced you to have sex with him when you didn’t want to? & 0 & 1 & 2 & 3\\
    12. &  used threats or intimidation to make you have sex when you did not want to? & 0 & 1 & 2 & 3\\
    13. &  used physical force or threats to make you do something else sexual that you did not want to do? & 0 & 1 & 2 & 3 \Bstrut\\
    \hline
    \end{tabular}
}
\end{table}

To analyze the data from the CTS, the analyst typically collapses an individual's responses to the violence items into a single binary measure that indicates whether the respondent reports any act of violence during the recall period. The resulting prevalence outcome is then used as the basis for statistical inference. In randomized evaluations, the prevalence outcomes in the treatment and control groups are compared to assess the effect of the program. 

We believe this coding strategy has become the norm in the field for several reasons. First, it was among the violence coding strategies suggested in \textcite{straus_measuring_1979} due to concerns about ``skewness'' in the distribution of responses on account of a minority of highly violent relationships. Second, a preference for binary coding is a historic legacy of the national and global prevalence surveys in the 1990s and 2000s where a prevalence measure was a natural summary statistic. Third, once adopted in seminal randomized evaluations it became important for subsequent studies to code their outcomes similarly for the purposes of comparability. However, a binary coding strategy discards information about frequency and severity, information which could be useful for providing a more nuanced understanding of program impacts. Interestingly, in the revised scale \textcite{straus_revised_1996} also suggested a summary measure they called \textit{chronicity} which was the sum of items scores among those reporting any violence, although this does not seem to have caught on. In this paper, we evaluate a binary coding strategy in terms of its efficiency --i.e. the consequences in terms of the statistical power and precision for answering substantive research questions-- and compare it to alternative strategies. 

\section{Theory} \label{sec:theory}

To better understand the implications of outcome coding choice in randomized evaluations, we first need a theoretical framework for describing how violence is distributed and how the effects of interventions operate. In this section, we develop a generative model for violence by returning to the original latent variable formulation of the CTS. We start with the simple case of defining a latent representation for a single act of violence and then move on to a model for multiple correlated acts which are more typical of how CTS-based instruments work. We then define causal effects on violence in the context of randomized evaluations where effects can vary across items and individuals. We conclude with the definition of several alternative coding strategies and a discussion of their theoretical advantages and disadvantages.

\subsection{Single act model}
\begin{figure}[t]
    \centering
    \begin{subfigure}[b]{0.49\textwidth}
    \centering
    \caption{Poisson}
    \includegraphics[width = \linewidth]{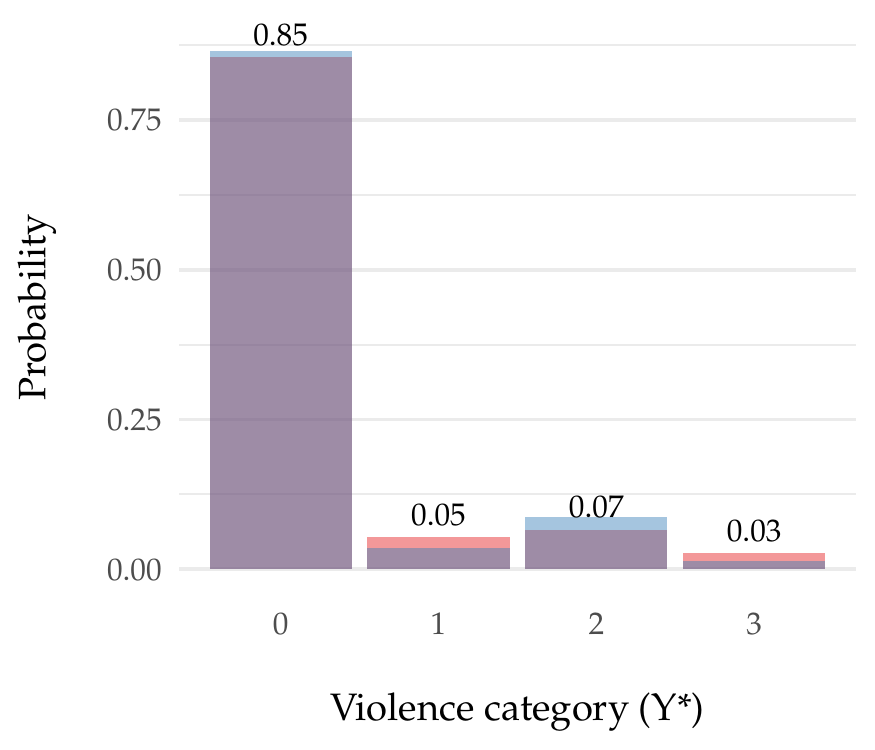}
    \end{subfigure}
    \begin{subfigure}[b]{0.49\textwidth}
    \centering
    \caption{Negative Binomial}
    \includegraphics[width = \linewidth]{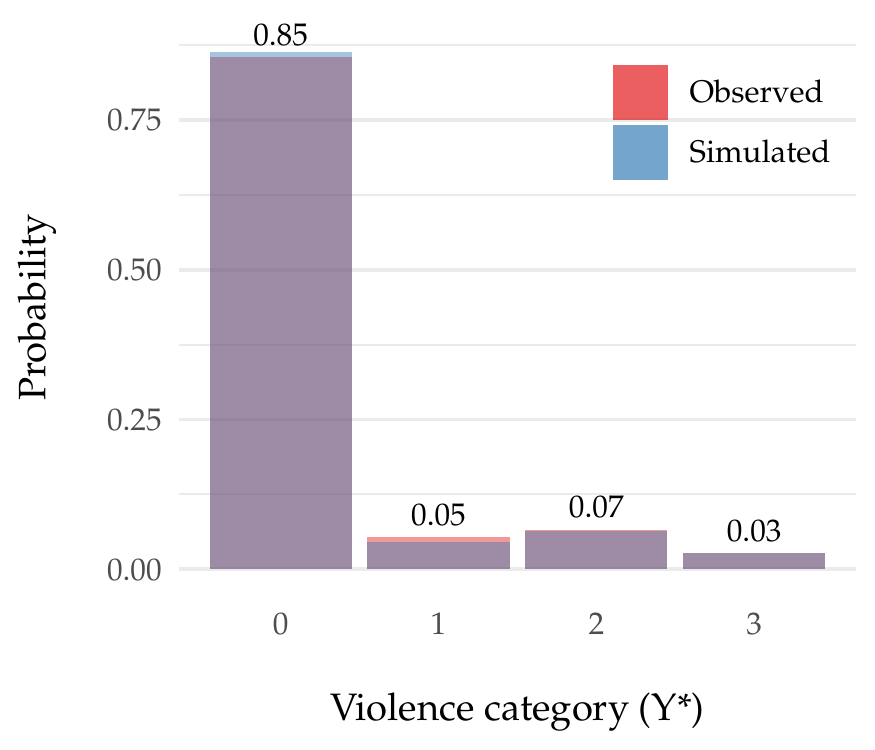}
    \end{subfigure}
    \caption{Distribution of observed acts vs. those simulated from (a) a zero-inflated Poisson with $\widehat{\lambda} = 2.36$ and $\widehat{\theta} = 0.84$ and (b) a zero-inflated Negative Binomial with $\widehat{\lambda} = 2.36$ and $\widehat{\theta} = 0.84$ based on maximum likelihood using data from Uganda.}
    \label{fig:single_act}
\end{figure}

Let $Y$ be the number of episodes of a specific violent act occurring over the follow up period, for instance the number of times the respondent is slapped. We assume that $Y$ is an i.i.d. realization from zero-inflated Poisson process of the form
\begin{equation*}
\begin{aligned}
Y  \sim \begin{cases} 0 & \text{with probability } \theta \\
   \text{Poisson}(\lambda) &\text{with probability } 1 - \theta
   \end{cases}
\end{aligned}
\end{equation*}
where the source of the excess zeros is a latent subpopulation of ``nonviolent'' couples. Here $\theta$ represents the probability that a woman is in a ``nonviolent'' relationship and $\lambda$ represents the average rate of violence among ``violent'' relationships.

In addition to excess zeros, it's possible that further heterogeneity exists between ``violent'' couples. This may be due to a long tail of couples in which acts occur more frequently. In this case, another possible generative model is a zero-inflated Negative Binomial, i.e.
\begin{equation*}
\begin{aligned}
Y  \sim \begin{cases} 0 & \text{with probability } \theta \\
   \text{NegBin}(\lambda, \phi) &\text{with probability } 1 - \theta
   \end{cases}
\end{aligned}
\end{equation*}
where the parameter $\phi$ captures the additional dispersion in violent acts.

As is common in the CTS, we assume that $Y$, the true number of acts, is further categorized at the time of survey measurement 
\[
Y^* = \begin{cases} 0 & \text{if }Y = 0 \\ 1 & \text{if } Y = 1 \\ 2 & \text{if } 2 \leq Y  \leq4 \\ 3 & \text{if } Y  \geq 5\end{cases}
\]
where 0 = ``Never'', 1 = ``Once'', 2 = ``A few times'', 3 = ``Many times''. In most surveys, we observe $Y^*$ while the true value $Y$ remains unknown. 

Figure \ref{fig:single_act} compares distribution of observed and simulated acts using data from on acts of slapping in a recent study in Uganda. We fit both Poisson and Negative Binomial models. Parameter values were estimated via maximum likelihood. We find that both models match the empirical data distribution quite well, but the more flexible Negative Binomial model better captures the distribution among violent couples.

\subsection{Multiple act model} \label{sec:multi_act}

In the CTS, violence is measured by multiple, potentially correlated, acts. We can generalize the single act model above by defining an i.i.d. vector $(Y_1, Y_2, \ldots, Y_{10})'$ where each element is now the number of reported acts of violence of each the ten types of acts given in Table 1. These are jointly distributed zero-inflated Poisson or zero-inflated Negative Binomial random variables

\[(Y_1, Y_2, \ldots, Y_{10})' \sim \text{ZIP}(\mathbf{\lambda}, \mathbf{\theta}, \mathbf{\Sigma}) \]
\[(Y_1, Y_2, \ldots, Y_{10})' \sim \text{ZINB}(\mathbf{\lambda}, \mathbf{\phi}, \mathbf{\theta}, \mathbf{\Sigma}) \]

where $\mathbf{\lambda}$, $\mathbf{\phi}$, and $\mathbf{\theta}$ are the vector analogs to $\lambda$, $\phi$, and $\theta$ defined previously but $\mathbf{\Sigma}$ is now a $10 \times 10$ variance-covariance matrix specifying the correlation structure between the types of violence. Figure \ref{fig:multi_act} below shows an example of the data generated in this case. In panel (c) the correlation matrix is typical of relationship between acts in many contexts. The highest correlations are between acts of sexual violence and acts of less extreme physical violence. In general, physical acts are more highly correlated with each other than with sexual acts and vice versa.

\begin{figure}[bp]
    \centering
    \begin{subfigure}[b]{0.9\textwidth}
    \centering
    \caption{Distribution of sum of $Y^*$}
    \includegraphics[width = \linewidth]{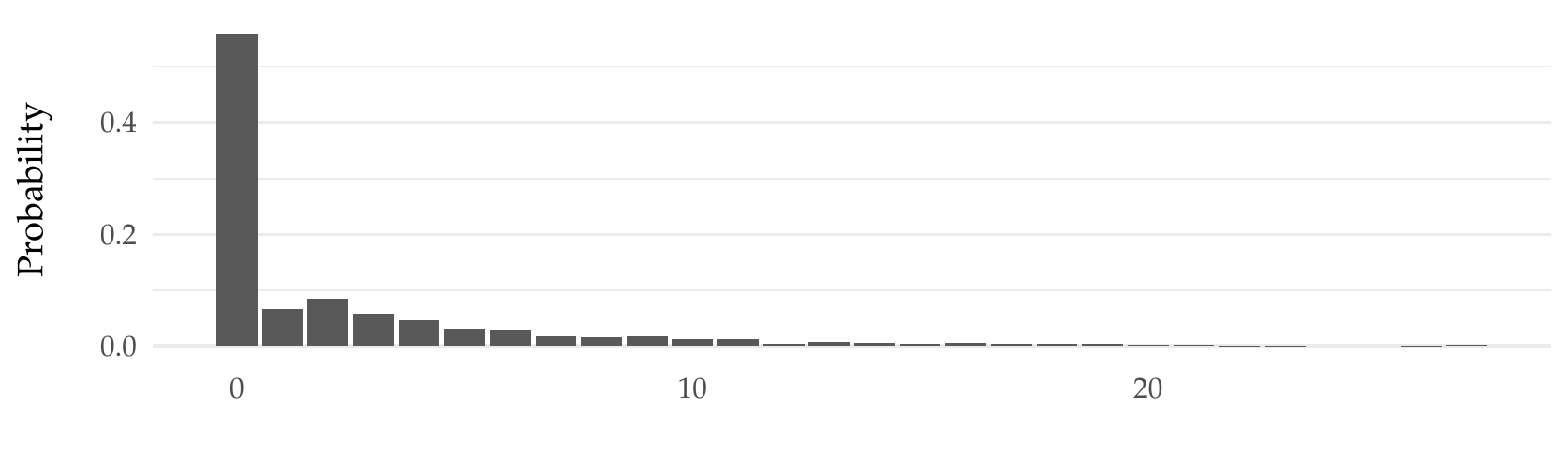}
    \end{subfigure}
    \begin{subfigure}[b]{0.9\textwidth}
    \centering
    \caption{Distribution of each act}
    \includegraphics[width = \linewidth]{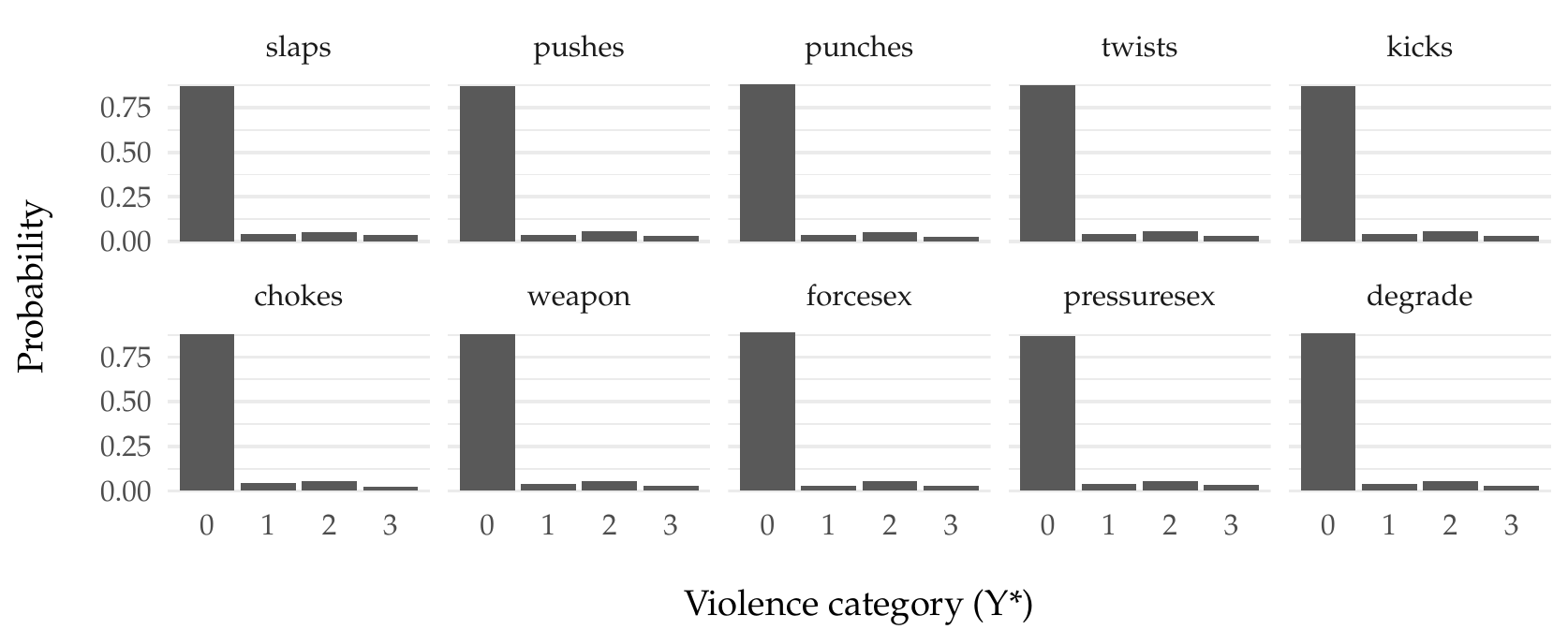}
    \end{subfigure}
    \begin{subfigure}[b]{0.9\textwidth}
    \centering
    \caption{Correlation}
    \includegraphics[width = \linewidth]{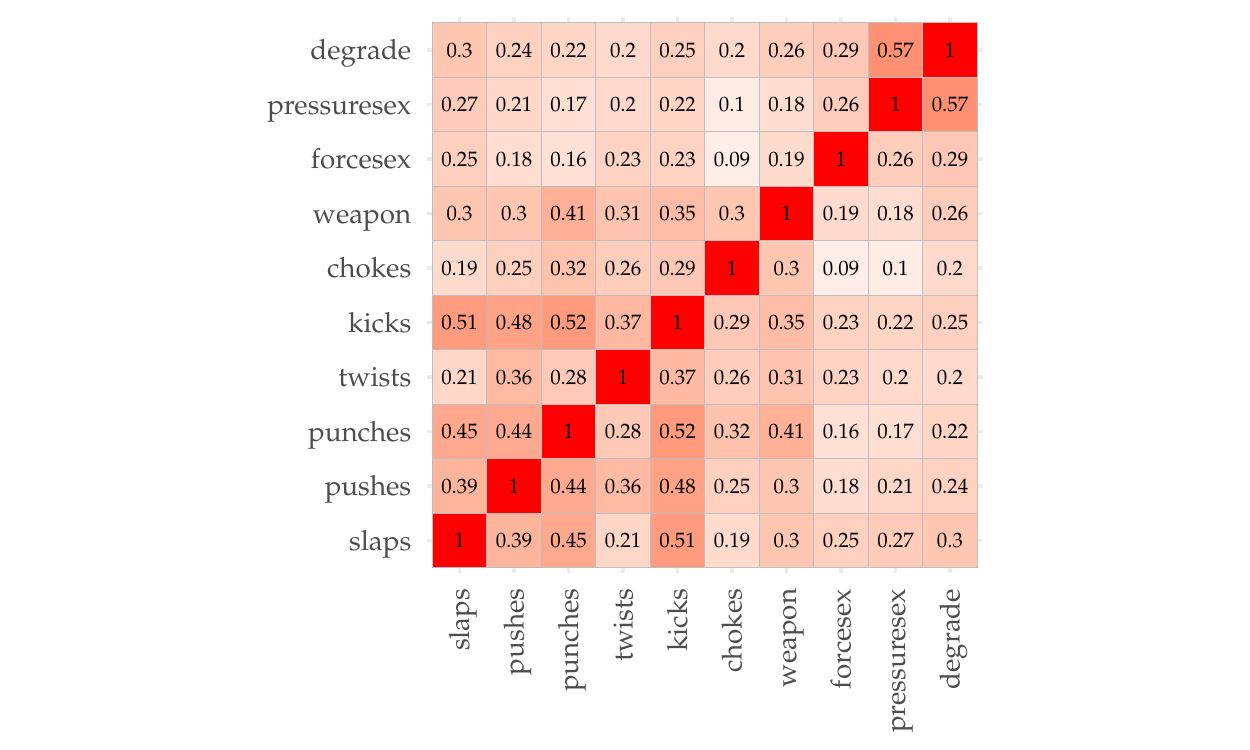}
    \end{subfigure}
    \caption{Distribution of multi-act Zero-inflated Poisson.}
    \label{fig:multi_act}
\end{figure}

\subsection{Intervention effects} \label{sec:po_model}
Now that we have a model for distribution of violence, we develop a framework for causal effects of an anti-violence program in a randomized experiment using the Neyman-Rubin causal model \cite{splawa-neyman_application_1990, rubin_estimating_1974}. Consider a collection of units $i = 1,\ldots,n$ assigned either to a hypothetical violence reduction program ($z = 1$) or control ($z = 0$). The objective is to measure latent violence $Y$ at a specific time after assignment of each unit. Let $Y_i(z)$ be a potential outcome representing the value of $Y$ if unit $i$ is assigned treatment $z$ for $z = 0, 1$, i.e. $Y_i(1)$ is the latent violence for unit $i$ when assigned to the violence reduction program and $Y_i(0)$ is the latent violence for unit $i$ when assigned to the control. Then the causal effect of the program for unit $i$ is defined as a comparison between the potential outcomes, e.g.
\[\tau_i = Y_i(1) - Y_i(0)\]
representing the counterfactual contrast within the same individual if given the program versus not. We can extend this individual effect to sample or population summaries such as the mean difference or ratio, e.g.
\[\operatorname{E}[\tau_i] = \operatorname{E}[Y_i(1)] - \operatorname{E}[Y_i(0)] \quad \text{or} \quad \operatorname{E}[\tau_i]  = \frac{\operatorname{E}[Y_i(1)]}{\operatorname{E}[Y_i(0)]}\]

In practice, we never observe both potential outcomes for any individual. Rather we assume we observe
\[Y_i = Z Y_i(1) + (1 - Z) Y_i(0)\]
where we observe $Y_i(1)$ for those assigned to treatment group and $Y_i(0)$ for those assigned to control. However, because randomization guarantees that distribution of potential outcomes will be independent of assignment $Z$, the average effect may be identified by a simple comparison of the observed outcomes in the treatment and control groups, e.g.
\[\operatorname{E}[\tau_i] = \operatorname{E}[Y_i \mid Z = 1] - \operatorname{E}[Y_i \mid Z = 0]\]

Based on our generative model above, we assume that violence under no intervention, $Y_i(0)$, follows a zero-inflated Poisson or Negative Binomial. To simulate effects in a trial we, also need to define possible changes in violence due to participation in an anti-violence program. Programs can affect violence in a variety of ways, they could have broad and consistent effects for all participants or only benefit a handful of the most engaged; they may influence certain types of violence or violent ``profiles'' more than others; they could have more mixed effects producing small improvements but also backlash; or they could prevent new cases of violence but leave those already experiencing violence without much benefit. Each of these in turn may have different implications for outcome coding choice.

While it is possible that program can lead to initiation of violence, we believe this is rare and therefore consider anyone who would not have experienced violence in absence of program remain violence free with program (i.e. $Y_i(1) = 0$ if $Y_i(0) = 0)$. For those who do experience violence in absence of program (i.e. $Y_i(0) > 0$), we assume that program effects fit into one of 4 possible response types ($S$): 
\begin{enumerate}
\item \textit{No effect} - the individual experiences the same violence regardless of whether they receive the program, i.e. $Y_i(1) = Y_i(0)$.
\item \textit{Cessation} - when exposed to the program, all violence stops regardless of frequency, i.e. $Y_i(1) = 0$ for any $Y_i(0)$.
\item \textit{Reduction} - when exposed to the program, violence is reduced by a fixed amount, i.e. $Y_i(1) < Y_i(0)$.
\item \textit{Increase} -  when exposed to the program, violence increases by a fixed amount, i.e. $Y_i(1) > Y_i(0)$.
\end{enumerate}
\begin{table}[t]
    \centering
    \caption{Example potential outcomes for possible response types}
    \label{tab:po_example}
    \begin{threeparttable}
    \renewcommand{\TPTminimum}{\linewidth}
    \makebox[\linewidth]{
    \begin{tabular}{cccccccc}    
    \toprule
     ID & type ($S$) & Z & $Y_i(1)$ & $Y_i(0)$ & $Y_i^*(1)$ & $Y_i^*(0)$ & $Y^*$ \\
     \midrule
     1  &   & 0 & 0 & 0 & 0  & 0 & 0 \\
     2  & 1 & 1 & 3 & 3 & 2  & 2 & 2 \\
     3  & 2 & 1 & 0 & 5 & 0  & 3 & 0 \\
     4  & 3 & 0 & 2 & 4 & 2  & 2 & 2 \\
     5  & 4 & 1 & 3 & 1 & 2  &1 & 2 \\
     \bottomrule \\
    \end{tabular}}
    \begin{tablenotes}[flushleft]
    \item \scriptsize{\textit{Notes:} Example potential outcome values for the response types defined in Section 2.3. $Y$ is true number of acts and $Y^*$ is categorization using WHO coding (i.e. 0 = None, 1 = Once, 2 = A few times, 3 = Many times). The first line shows an individual who experiences no violence in absence of program and therefore is assumed to also experience no violence when assigned to program $Y_i(1) = Y_i(0) = 0$. The second line is a type 1 individual for whom the program has no effect, i.e. the level of violence they experience when given the program is the same as when not given $Y_i(1) = Y_i(0) = 3$. The third line is a type 2 individual for whom violence ceases when given the program $Y_i(1) = 0$. The fourth line is a type 3 individual for whom violence is reduced when given the program $Y_i(1) = 2 < Y_i(0) = 4$. Finally, the last line is a type 4 individual for whom violence increases as a results of the program $Y_i(1) = 3 > Y_i(0) = 1$.}
\end{tablenotes}
\end{threeparttable}
\end{table}
Table \ref{tab:po_example} shows example values of $Y_i(1)$, $Y_i(0)$ as well as their survey-measured equivalents for each response type. The response types allow us to specify a variety of program effects in terms of the relative frequencies of the response types. For example, in a population where background violence prevalence is 40\%, among those with violence a particular program may have no effect for 50\%, lead to cessation entirely for 20\%, reduction, but not cessation for 20\%, and an increase for 10\%.

In practice, we draw response types for violent couples from a multinomial distribution
\[S \sim \operatorname{Multinomial}(p_s)\]
where $p_s$ is a length 4 vector of relative proportions of each response type, then given $S$ the $Y_i(1)$ for each individual can be determined from

\[Y_i(1) = \begin{cases} 0 & \text{if } Y_i(0) = 0 \\ Y_i(0) & \text{if } S = 1 \text{ and } Y_i(0) > 0 \\ 0 & \text{if } S = 2 \text{ and } Y_i(0) > 0 \\ Y_i(0) - x & \text{if } S = 3 \text{ and } Y_i(0) > 0 \\ Y_i(0) + x & \text{if } S = 4 \text{ and } Y_i(0) > 0 \end{cases}.
\]

Finally, we also allow for the possibility that programs affect acts of violence differently. For instance a program may lead to reductions and/or cessations of moderate acts only or a consent-based program may affect sexual acts while leaving physical acts largely unchanged. 

\subsection{Outcome coding strategies}
Analysts typically collapse responses to multiple acts into a single summary measure of violence. However, there are many possible strategies for doing so. Here, we consider two common outcome coding strategies. The first is based on the strategy discussed in section \ref{sec:cts} which collapses the items in the CTS scale into a single binary measure representing whether the woman reported any acts of violence during the recall period. 
\[Y^*_{binary} = \begin{cases} 0 & \text{if all } Y^*_1 = 0, Y^*_2 = 0, \ldots, Y^*_{K} = 0  \\ 1 & \text{if any } Y^*_1 > 0 \text{ or } Y^*_2 > 0 \text{ or }  \ldots \text{ or } Y^*_{K} > 0 \end{cases}\]
Treatment effects based on this outcome represent difference in probability of reporting any violence. Depending on prior history, observed changes in this measure may consists of cessation of ongoing violence or prevention of new cases of violence or a combination of the two. 

The second outcome coding strategy is a continuous measure that is still straightforward to construct, but less common: a simple sum of the $K$ items. 
\[Y^*_{sum} = Y^*_1 + Y^*_2 + \ldots + Y^*_{K}\]
Treatment effects based on this outcome represent differences in number of acts of violence if the true number of acts is recorded, but is a little more difficult to interpret if the CTS categories are used, i.e. 0 = ``Never'', 1 = ``Once'', 2 = ``A few times'', 3 = ``Many times''. It treats all acts as essentially the same (regardless of severity), but can reflect greater gradation in the amount of violence reported. In simulations, we divide $Y^*_{sum}$ by the total possible score to normalize values between 0 and 1 and to make variance comparisons easier as both $Y^*_{binary}$ and $Y^*_{sum}$ are then on the same scale.

\subsection{Estimation}
For both continous and binary outcomes, we use a least squares regression estimator to estimate the average treatment effect of the form
\[Y_i = \alpha + \tau Z_i + \varepsilon_i\]
where $Z_i$ is the random assignment indicator and $\tau$ is the average treatment effect. We calculate robust standard errors using ``HC2'' formulation from \texttt{estimatr} package in R. We could use a nonlinear model such as logit or probit for the binary outcome or a poisson or negative binomial for the continuous, however we don't because (1) we are principally concerned with coding rather than estimation, (2) least squares with robust SEs is still unbiased, and (3) often the difference in average partial effects is quite small.

\section{Simulation} \label{sec:data}

To explore the effects of outcome coding choice on estimation, we conduct a series of finite sample Monte Carlo simulations using the \href{https://declaredesign.orgg}{DeclareDesign} package in R. We generate potential outcome data according to the model defined above and set values of $\mathbf{\lambda}$, $\mathbf{\phi}$, and $\mathbf{\theta}$ based on empirical estimates or draw directly from the empirical distributions. In all simulations we generate a full set of potential outcomes for a sample of size $N$, randomize half to a hypothetical program and half to control, apply outcome coding choices, and estimate program effects. We repeat this 1000 times and calculate the following performance statistics:

\begin{itemize}
    \item \textit{Bias}: average difference between estimate $\hat{\tau}_m$ in each simulation and the true value $\tau$.
    \[\frac{1}{M} \sum_{m=1}^M(\hat{\tau}_m - \tau)\]
    \item \textit{Root-mean-square error (RMSE)}: square root of the average squared distance between estimate $\hat{\tau}_m$ in each simulation and the true value $\tau$.
    \[\sqrt{\frac{1}{M}\sum_{m=1}^M(\hat{\tau}_m - \tau)^2}\]
    \item \textit{Power}: in a null hypothesis significance testing framework, the probability of correctly rejecting the null when the null is false. 
    \item \textit{Coverage}: The proportion of simulations in which the confidence interval for $\hat{\tau}_m$ contains the true value $\tau$.
\end{itemize}

We conduct a series of experiments in which we vary the treatment effect structures and then compare different outcome coding choices under consistent estimation strategies. Building on our potential outcomes model in \ref{sec:po_model}, we consider four possible violence reduction scenarios consisting of different assumed proportions of response types: (1) \textit{cessation only} - violence ceases for 30\% of individuals and there is no effect for remaining 70\%, (2)  \textit{cessation + reduction} - violence ceases for 10\% of individuals, is reduced but not ceased for 20\% and 70\% are unaffected, (3) \textit{reduction only} - violence reduces, but does not cease, for 30\% of individuals and 70\% are unaffected, (4) \textit{cessation + reduction + increase} - violence ceases for 10\%, reduces, but does not cease, for 15\%, increases for 5\% and 70\% are unaffected. In all scenarios we assume 70\% of individuals are unaffected, which may seem high, but consider that most trials are powered for to detect smaller effects than this. For example for the cessation scenario, the 30\% affected translates to a risk ratio of 0.7 for the binary measure. Finally, we also vary whether reductions affect all acts equally or only a subset of acts. 

\begin{table}

\caption{Simulation results based on Becoming One study in Uganda.\label{tab:b1_sims}}
\centering
\fontsize{7}{9}\selectfont
\begin{threeparttable}
\begin{tabular}[t]{lcccccccccccc}
\toprule
\multicolumn{1}{c}{ } & \multicolumn{4}{c}{All acts} & \multicolumn{4}{c}{Physical only} & \multicolumn{4}{c}{Sexual only} \\
\cmidrule(l{3pt}r{3pt}){2-5} \cmidrule(l{3pt}r{3pt}){6-9} \cmidrule(l{3pt}r{3pt}){10-13}
 & Bias & RMSE & Power & Coverage & Bias & RMSE & Power & Coverage & Bias & RMSE & Power & Coverage\\
\midrule
\addlinespace[0.3em]
\multicolumn{13}{l}{\textbf{Cessation only}}\\
\hspace{1em}$Y_{binary}$ & 0 & 0.020 & 1.000 & 0.957 & 0 & 0.022 & 0.289 & 0.946 & 0 & 0.021 & 0.682 & 0.956\\
\hspace{1em}$Y_{sum}$ & 0 & 0.006 & 0.972 & 0.964 & 0 & 0.006 & 0.511 & 0.957 & 0 & 0.006 & 0.375 & 0.959\\
\addlinespace[0.3em]
\multicolumn{13}{l}{\textbf{Cessation + reduction}}\\
\hspace{1em}$Y_{binary}$ & 0 & 0.021 & 0.488 & 0.949 & 0 & 0.022 & 0.071 & 0.953 & 0 & 0.022 & 0.125 & 0.953\\
\hspace{1em}$Y_{sum}$ & 0 & 0.006 & 0.551 & 0.961 & 0 & 0.006 & 0.194 & 0.960 & 0 & 0.006 & 0.152 & 0.955\\
\addlinespace[0.3em]
\multicolumn{13}{l}{\textbf{Reduction only}}\\
\hspace{1em}$Y_{binary}$ & 0 & 0.022 & 0.048 & 0.952 & 0 & 0.022 & 0.048 & 0.952 & 0 & 0.022 & 0.048 & 0.952\\
\hspace{1em}$Y_{sum}$ & 0 & 0.006 & 0.255 & 0.957 & 0 & 0.006 & 0.103 & 0.954 & 0 & 0.006 & 0.096 & 0.957\\
\addlinespace[0.3em]
\multicolumn{13}{l}{\textbf{Cessation + reduction + increase}}\\
\hspace{1em}$Y_{binary}$ & 0 & 0.021 & 0.481 & 0.953 & 0 & 0.022 & 0.072 & 0.949 & 0 & 0.022 & 0.127 & 0.955\\
\hspace{1em}$Y_{sum}$ & 0 & 0.006 & 0.350 & 0.957 & 0 & 0.006 & 0.136 & 0.956 & 0 & 0.006 & 0.113 & 0.953\\
\bottomrule
\end{tabular}
\begin{tablenotes}[para]
\item \textit{Notes:} 
\item Monte carlo performance estimates based on 1,000 simulated assignments and 100 bootstrap samples. 
\end{tablenotes}
\end{threeparttable}
\end{table}

Table \ref{tab:b1_sims} shows the results for simulations on the multiple act model in section \ref{sec:multi_act} based on empirical distribution of the Becoming One (B1) trial in Uganda \cite{boyer_religious_2022}. Both the sum and binary measures are unbiased and demonstrate good coverage for their respective estimands across all scenarios when we use the CTS coding as ``truth'' and ignore the latent true number of acts. If we do consider the latent number of acts then bias and poor coverage is possible. The RMSE is also always lower for the continuous measure as compared to the binary (after dividing by 30 to convert to similar range), reflecting less variability in estimates from simulation to simulation. For power the results are a bit more mixed, the binary measure is higher powered when cessation (or no effect) is the only possible effect of the program. This makes some intuitive sense as the extra information provided by the continuous measure is irrelevant if effects are either all or nothing. When there is some portion of the sample for whom violence is reduced, but does not cease, and no one's violence is increased by the program the continuous sum is better powered. When all response types are possible either measure can be higher powered.

\begin{figure}[t]
 \centering
 \includegraphics[width=\textwidth]{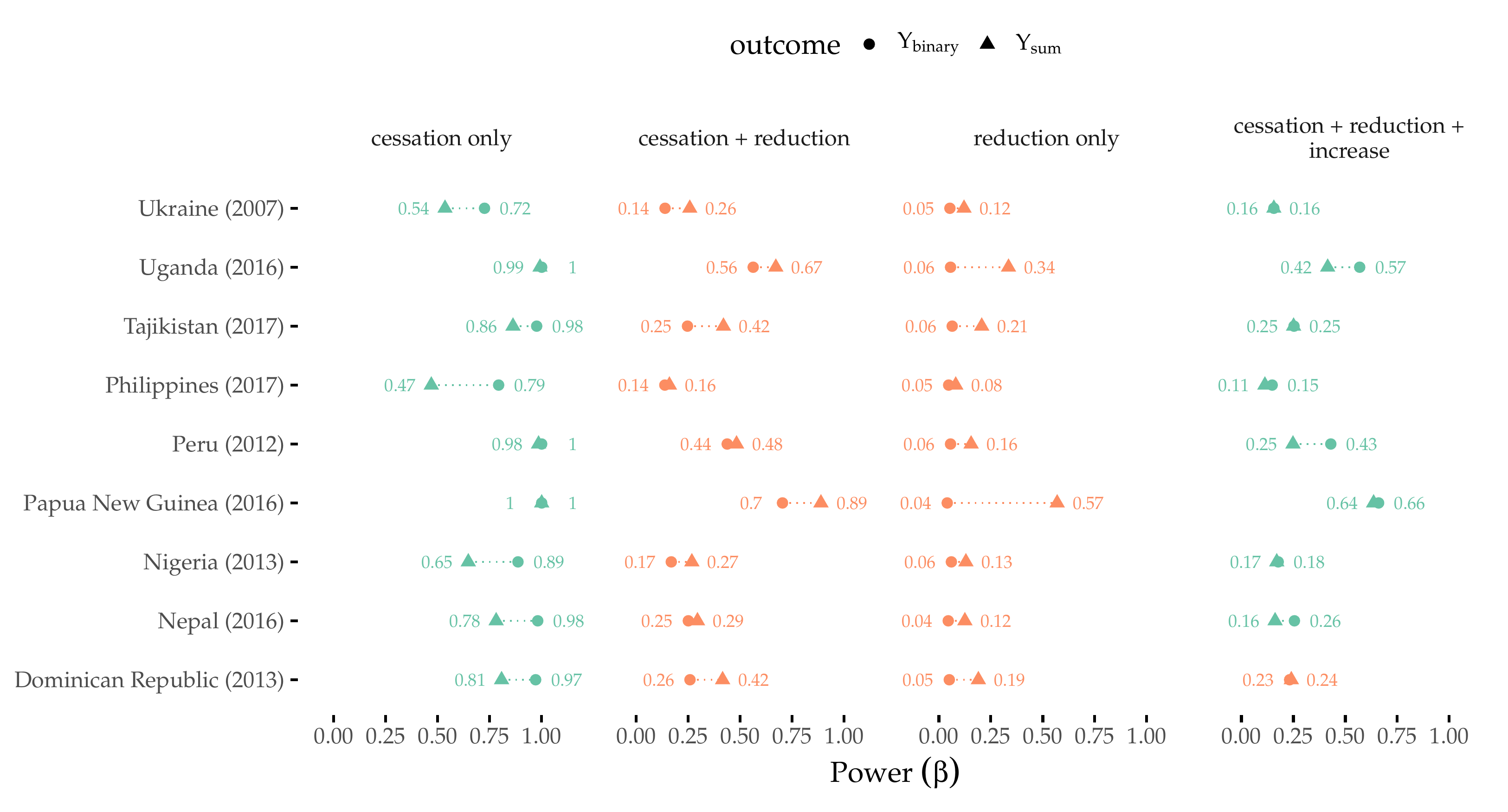}
 \caption{Simulated power differences between binary and sum outcome codings across contexts.}
 \label{fig:dhs_results}
\end{figure}

To determine whether these findings are affected by different untreated distributions of violence, we examined additional $Y(0)$ outcome distributions based on empirical estimates in a variety of settings from the Demographic and Health Surveys. We chose nine representative countries, with 12-month prevalences ranging from 5.4\% in the Philippines to 47.6\% in Papua New Guinea. Figure \ref{fig:dists} in the appendix plots the distributions for each act in each country. In addition to variability in overall prevalence, these settings reflect heterogeneity in types of violence which predominate, with some countries having significantly more sexual violence than others or differences in distribution of act frequencies. Figure \ref{fig:dhs_results} plots the differences in power between binary and continuous sum. Despite the heterogeneity in prevalence and distribution of untreated violence, the findings are broadly similar. When the primary action is cessation, the binary measure is generally more powerful. When there is reduction but no increase the continuous sum is higher powered. When there is a subset for whom violence increases results are mixed, although in these nine examples the binary is often higher powered. In appendix Table \ref{tab:dhs_sims_power}, we include additional simulation results in which effects are concentrated on only physical, only sexual, or only moderate acts. When only physical acts are affected by the program, results are largely unchanged from above. When only sexual acts are affected by the program, the continuous sum measure now dominates the binary even when cessation is the only action. This is at least partly because there are only 3 acts of sexual violence (compared to 7 for physical violence) in the CTS-based scale, and these acts are correlated with physical acts. When only moderate acts are affected by the program, the continuous measure often, but not always is higher powered. Again, this depends on whether there are a subset of couples who only experience more moderate acts and how large this subset is relative to the other as well as how often those who experience normatively more severe acts also experience moderate ones. 

\section{Application} \label{sec:application}
In this section, we assemble data from recent trials of anti-violence programs and examine whether outcome coding choice materially affected interpretation of program impacts. Seven trials contributed individual data which were re-analyzed: Bandebereho \cite{doyle_gender-transformative_2018} , Becoming One \cite{boyer_religious_2022} (Uganda), Indashyikirwa \cite{dunkle_effective_2020} (Rwanda), MAISHA CRT01 \cite{kapiga_social_2019} (Tanzania), MAISHA CRT02 (Tanzania), Stepping Stones \cite{jewkes_impact_2008} (South Africa), and Unite for Better Life \cite{sharma_effectiveness_2020} (Ethiopia). 

In most trials the binary and continuous measures lead to similar conclusions. Table \ref{tab:application} highlights notable discrepancies from two trials: Becoming One (B1) and MAISHA CRT01. In the case of the former, at the 6 month follow up, a non-significant 2.6 percentage point reduction in the binary measure of violence was observed ($p = 0.265$). However, the reduction in the continuous sum measure of violence was significant ($p = 0.014$). Standard errors for each additionally suggested the greater precision of the continuous measure. In exploratory analyses, comparisons of the underlying distributions (Figure \ref{fig:b1_dist}) suggested that the additional reductions, but not cessations, in the tail of the distribution drove differences in the two measures. At endline, 12 months after the start of the program, both measures showed significant reductions. This was consistent with the hypothesis that changes in relationship dynamics took time to occur as couples engaged with the program.

In the MAISHA trial, we see the opposite: a non-significant (at conventional pre-specified $\alpha$ level of 0.05) but precisely estimated reduction in the binary measure ($p = 0.051$) but a more imprecise non-significant decline in the continuous sum measure. This discrepancy could be consistent with either: (a) cessation being the primary mechanism of changes in violence due to the program or (b) a small subpopulation of already violent couples for whom the program may have caused increases in violence which is picked up by the continuous but not the binary measure. Given the sample size, it could always of course also be noise. Additional, exploratory analyses can help reveal what is more likely. However, comparing measures can have important implications for how trial results are interpreted 

\begin{table}[t]
    \centering
    \caption{Example discrepancies in results by outcome coding choice in recent trials    \label{tab:application}}
    \begin{threeparttable}
    
    \begin{tabular}{lccccccccc}
    \toprule
     & & \multicolumn{4}{c}{$Y_{binary}$} & \multicolumn{4}{c}{$Y_{sum}$} \\
     \cmidrule(l{3pt}r{3pt}){3-6}  \cmidrule(l{3pt}r{3pt}){7-10}
     Trial & N &  T & C & Diff & $p$-value & T & C & Diff & $p$-value \\
    \midrule
    B1\tnote{a} & 1,680 & 35.1\% & 37.7\% & -2.6\% & 0.265 & 1.23 & 1.61 & -0.38 & 0.014 \\ 
    MAISHA 1\tnote{b} & 919 & 23.1\% & 27.4\% & -4.3\% & 0.051 & 1.39 & 1.49 & -0.10 & 0.266 \\
    \bottomrule
    \end{tabular}
    \begin{tablenotes}[para]
\item[a] Midline (6 month) results from Becoming One study in Uganda.
\item[b] Endline (24 month) results from MAISHA trial I in Tanzania. \\
\end{tablenotes}
\end{threeparttable}
\end{table}

\section{Discussion} \label{sec:discussion}

In this study, we developed a generative model for violence and violence reduction in trials and used it to better understand how outcome coding choice affects statistical efficiency. When comparing two simple measures: a binary indicator for whether any act of violence was reported or a continuous sum of act frequency categories. We find that no measure strictly dominates, as there are settings where the binary measure may be preferred, however the continuous measure was higher powered in the majority of scenarios considered. We therefore recommend that trialists report both measures, when possible in order to better facilitate interpretation, particularly in circumstances such as those highlighted in section \ref{sec:application} where there are discrepancies.  

We also encourage trialists to consider more detailed power simluations, such as those conducted here, when planning future trials. To make them as relevant as possible, these can be based on empirical distributions previously collected data or from baseline. To assist in this we have made our code freely available at \href{https://github.com/boyercb/ipv-measurement}{https://github.com/boyercb/ipv-measurement} to practitioners. 

Our main finding contrasts somewhat with the literature on ``dichotomania'' in the medical sciences \cite{senn_dichotomania_2005, senn_measurement_2009}, which holds that dichotomization of trial outcomes is often a scourge to be avoided, principally because dichotomous outcomes are less efficient. This is often based on theory which shows that the dichotomization of a single normally distribution outcome leads to loss of efficiency. However, the violence setting is unique. As developed in section \ref{sec:theory}, there are clear theoretical reasons why the effect of the program itself may be dichotomous for a subset or even a majority of people. Further, heterogeneity in effects may also lead to sub-populations for whom the program increases violence. This may be particularly true in settings where backlash is possible. 

Researchers may have reasons beyond efficiency for preferring one measure over another. In some settings, violence cessation may be the only substantively interesting or meaningful effect. In other settings, one measure may be perceived as more reliable than another. In this study, for clarity, we chose to focus on only a small subset of the trade-offs that trialists must make. However, our recommendation is that, when possible, trialists should report multiple measures that can still aid in interpretation of findings.

Our study contributes to a larger literature about the measurement of violence both within the context of randomized trials as well as more broadly. Several authors have focused on the notable limitation that most violence is self-reported \cite{cullen_method_2020, park_private_2021, peterman_list_2018,stark_disclosure_2017,gibson_measuring_2022}, which is a particular concern in non-blinded randomized trials where certain incentives may drive differential reporting. Some have proposed alternative strategies like self-administration, list-randomization, or randomized response that confer a greater degree of anonymity. The present paper sidesteps this concern by focusing instead on the statistical properties of outcome coding choices. While we use the standard self-reported modules as the basis for our model, our results are agnostic to how violence is measured and could equally be applied to violence as assessed via different means. However, as previous commentators have noted several of these anonymized alternatives rely on asymptotic comparisons or marginal differences in responses and thereby sacrifice a considerable amount of statistical efficiency.

An additional value of our study is that our ``first-principles'' approach can be applied to assist in other pertinent measurement questions in violence research. For instance, for some interventions to reduce violence there is a theoretical question as to whether ``backlash'' \cite{chin_male_2012} might ensue for some couples. Our approach makes it easy to simulate different ``backlash'' structures in a variety of settings and can assist in figuring out the optimal statistical procedure for estimating ``backlash''-type effects when they do exist.

\clearpage

\singlespacing
\printbibliography

\clearpage

\section*{Appendix A.} \label{sec:appendixa}
\addcontentsline{toc}{section}{Appendix A}
\setcounter{figure}{0}
\renewcommand\thefigure{A\arabic{figure}}    
\setcounter{table}{0}
\renewcommand\thetable{A\arabic{table}}

\begin{figure}[p]
 \centering
 \includegraphics[width=\textwidth]{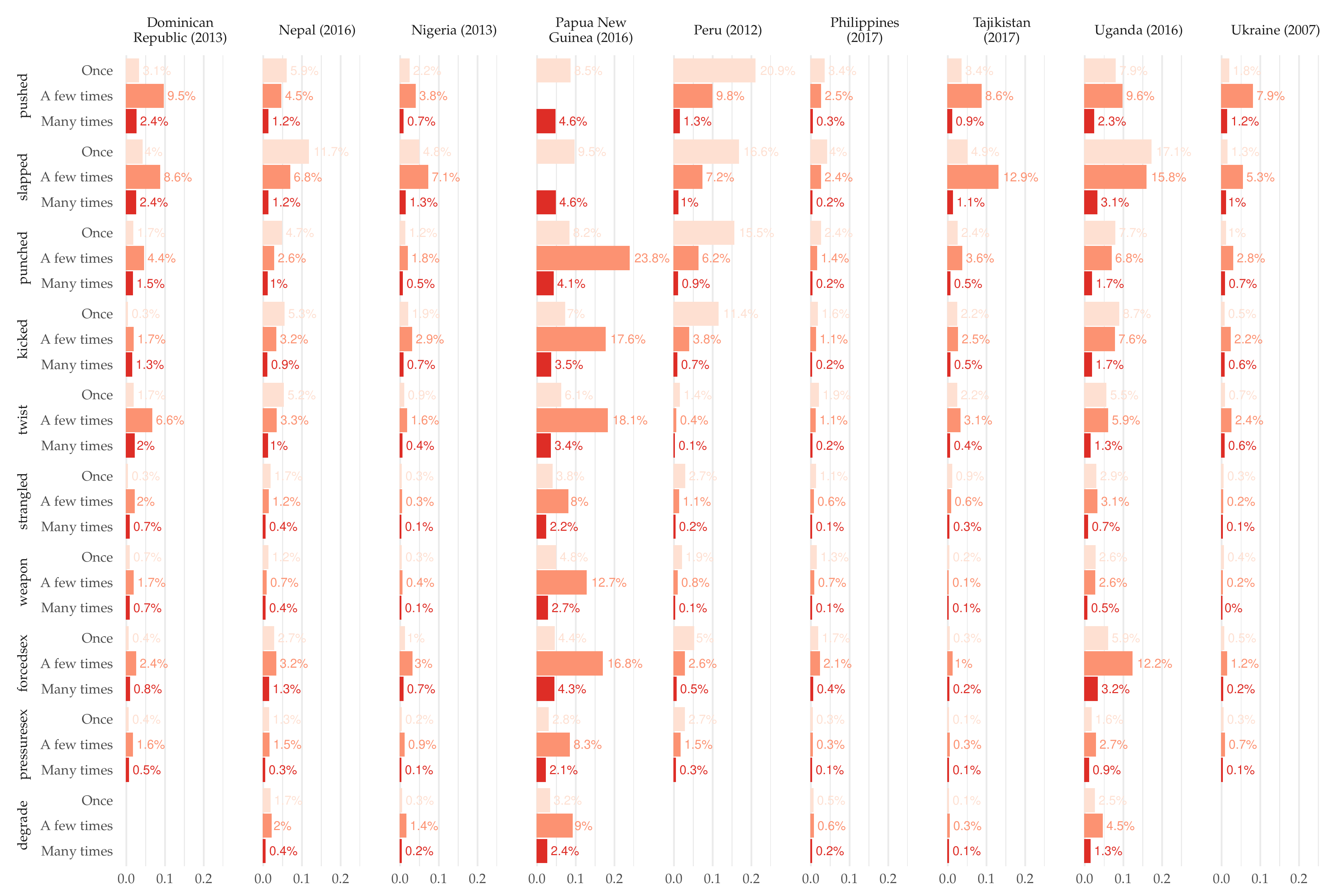}
 \caption{Empirical distributions of violent acts from select Demographic and Health Surveys.}
 \label{fig:dists}
\end{figure}

\begin{landscape}
\begin{table}[b]

\caption{Full results for monte carlo simulation of power using DHS data.\label{tab:dhs_sims_power}}
\centering
\fontsize{7}{9}\selectfont
\begin{threeparttable}
\begin{tabular}[t]{lp{0.6cm}p{0.6cm}p{0.6cm}p{0.6cm}p{0.6cm}p{0.6cm}p{0.6cm}p{0.6cm}p{0.6cm}p{0.6cm}p{0.6cm}p{0.6cm}p{0.6cm}p{0.6cm}p{0.6cm}p{0.6cm}p{0.6cm}p{0.6cm}}
\toprule
\multicolumn{1}{c}{ } & \multicolumn{2}{b{1.4cm}}{\centering Ukraine (2007)} & \multicolumn{2}{b{1.4cm}}{\centering Nepal (2016)} & \multicolumn{2}{b{1.4cm}}{\centering Philippines (2017)} & \multicolumn{2}{b{1.4cm}}{\centering Peru (2012)} & \multicolumn{2}{b{1.4cm}}{\centering Dominican Republic (2013)} & \multicolumn{2}{b{1.4cm}}{\centering Tajikistan (2017)} & \multicolumn{2}{b{1.4cm}}{\centering Uganda (2016)} & \multicolumn{2}{b{1.4cm}}{\centering Nigeria (2013)} & \multicolumn{2}{b{1.4cm}}{\centering Papua New Guinea (2016)} \\
\cmidrule(l{3pt}r{3pt}){2-3} \cmidrule(l{3pt}r{3pt}){4-5} \cmidrule(l{3pt}r{3pt}){6-7} \cmidrule(l{3pt}r{3pt}){8-9} \cmidrule(l{3pt}r{3pt}){10-11} \cmidrule(l{3pt}r{3pt}){12-13} \cmidrule(l{3pt}r{3pt}){14-15} \cmidrule(l{3pt}r{3pt}){16-17} \cmidrule(l{3pt}r{3pt}){18-19}
 & $Y_{binary}$ & $Y_{sum}$ & $Y_{binary}$ & $Y_{sum}$ & $Y_{binary}$ & $Y_{sum}$ & $Y_{binary}$ & $Y_{sum}$ & $Y_{binary}$ & $Y_{sum}$ & $Y_{binary}$ & $Y_{sum}$ & $Y_{binary}$ & $Y_{sum}$ & $Y_{binary}$ & $Y_{sum}$ & $Y_{binary}$ & $Y_{sum}$\\
\midrule
\addlinespace[0.3em]
\multicolumn{19}{l}{\textbf{All acts}}\\
\hspace{1em}cessation only & 0.971 & 0.807 & 0.981 & 0.781 & 0.886 & 0.647 & 1.000 & 1.000 & 1.000 & 0.985 & 0.793 & 0.469 & 0.976 & 0.862 & 1.000 & 0.992 & 0.725 & 0.535\\
\hspace{1em}cessation + reduction & 0.258 & 0.415 & 0.250 & 0.294 & 0.168 & 0.267 & 0.703 & 0.888 & 0.437 & 0.482 & 0.137 & 0.159 & 0.246 & 0.419 & 0.562 & 0.671 & 0.138 & 0.257\\
\hspace{1em}reduction only & 0.050 & 0.190 & 0.045 & 0.125 & 0.060 & 0.130 & 0.040 & 0.569 & 0.056 & 0.155 & 0.047 & 0.081 & 0.064 & 0.206 & 0.056 & 0.335 & 0.053 & 0.121\\
\hspace{1em}cessation + reduction + increase & 0.232 & 0.240 & 0.255 & 0.162 & 0.177 & 0.170 & 0.660 & 0.636 & 0.430 & 0.248 & 0.148 & 0.113 & 0.253 & 0.251 & 0.569 & 0.415 & 0.156 & 0.156\\
\addlinespace[0.3em]
\multicolumn{19}{l}{\textbf{Physical only}}\\
\hspace{1em}cessation only & 0.903 & 0.705 & 0.740 & 0.573 & 0.577 & 0.456 & 0.972 & 0.984 & 0.988 & 0.937 & 0.426 & 0.310 & 0.955 & 0.801 & 0.922 & 0.917 & 0.596 & 0.489\\
\hspace{1em}cessation + reduction & 0.182 & 0.321 & 0.138 & 0.191 & 0.102 & 0.163 & 0.232 & 0.720 & 0.306 & 0.398 & 0.090 & 0.129 & 0.212 & 0.360 & 0.204 & 0.434 & 0.102 & 0.204\\
\hspace{1em}reduction only & 0.044 & 0.161 & 0.042 & 0.089 & 0.050 & 0.098 & 0.060 & 0.380 & 0.065 & 0.126 & 0.060 & 0.079 & 0.061 & 0.209 & 0.040 & 0.162 & 0.040 & 0.124\\
\hspace{1em}cessation + reduction + increase & 0.186 & 0.213 & 0.121 & 0.115 & 0.117 & 0.121 & 0.253 & 0.436 & 0.290 & 0.211 & 0.077 & 0.088 & 0.208 & 0.231 & 0.203 & 0.233 & 0.090 & 0.126\\
\addlinespace[0.3em]
\multicolumn{19}{l}{\textbf{Sexual only}}\\
\hspace{1em}cessation only & 0.041 & 0.058 & 0.050 & 0.080 & 0.069 & 0.066 & 0.065 & 0.191 & 0.047 & 0.070 & 0.049 & 0.062 & 0.046 & 0.060 & 0.124 & 0.198 & 0.045 & 0.050\\
\hspace{1em}cessation + reduction & 0.046 & 0.043 & 0.056 & 0.063 & 0.049 & 0.071 & 0.058 & 0.099 & 0.052 & 0.057 & 0.054 & 0.058 & 0.053 & 0.046 & 0.067 & 0.109 & 0.048 & 0.053\\
\hspace{1em}reduction only & 0.045 & 0.044 & 0.055 & 0.052 & 0.052 & 0.050 & 0.060 & 0.075 & 0.058 & 0.059 & 0.049 & 0.045 & 0.071 & 0.052 & 0.062 & 0.083 & 0.052 & 0.053\\
\hspace{1em}cessation + reduction + increase & 0.040 & 0.057 & 0.050 & 0.051 & 0.046 & 0.052 & 0.043 & 0.081 & 0.052 & 0.045 & 0.059 & 0.057 & 0.045 & 0.055 & 0.064 & 0.084 & 0.056 & 0.051\\
\addlinespace[0.3em]
\multicolumn{19}{l}{\textbf{Moderate only}}\\
\hspace{1em}cessation only & 0.227 & 0.210 & 0.230 & 0.175 & 0.176 & 0.167 & 0.203 & 0.383 & 0.376 & 0.482 & 0.143 & 0.117 & 0.535 & 0.433 & 0.256 & 0.346 & 0.195 & 0.204\\
\hspace{1em}cessation + reduction & 0.055 & 0.104 & 0.072 & 0.071 & 0.071 & 0.099 & 0.056 & 0.189 & 0.063 & 0.140 & 0.065 & 0.068 & 0.088 & 0.171 & 0.066 & 0.128 & 0.078 & 0.104\\
\hspace{1em}reduction only & 0.060 & 0.091 & 0.048 & 0.054 & 0.054 & 0.069 & 0.052 & 0.089 & 0.053 & 0.071 & 0.063 & 0.051 & 0.058 & 0.126 & 0.053 & 0.069 & 0.054 & 0.073\\
\hspace{1em}cessation + reduction + increase & 0.076 & 0.095 & 0.060 & 0.066 & 0.063 & 0.073 & 0.072 & 0.111 & 0.078 & 0.093 & 0.048 & 0.055 & 0.106 & 0.135 & 0.063 & 0.102 & 0.058 & 0.082\\
\bottomrule
\end{tabular}
\begin{tablenotes}[para]
\item \textit{Notes:} 
\item Monte carlo performance estimates based on 1,000 simulated assignments and 100 bootstrap samples. 
\end{tablenotes}
\end{threeparttable}
\end{table}

\end{landscape}

\begin{figure}[p]
 \centering
 \includegraphics[width=\textwidth]{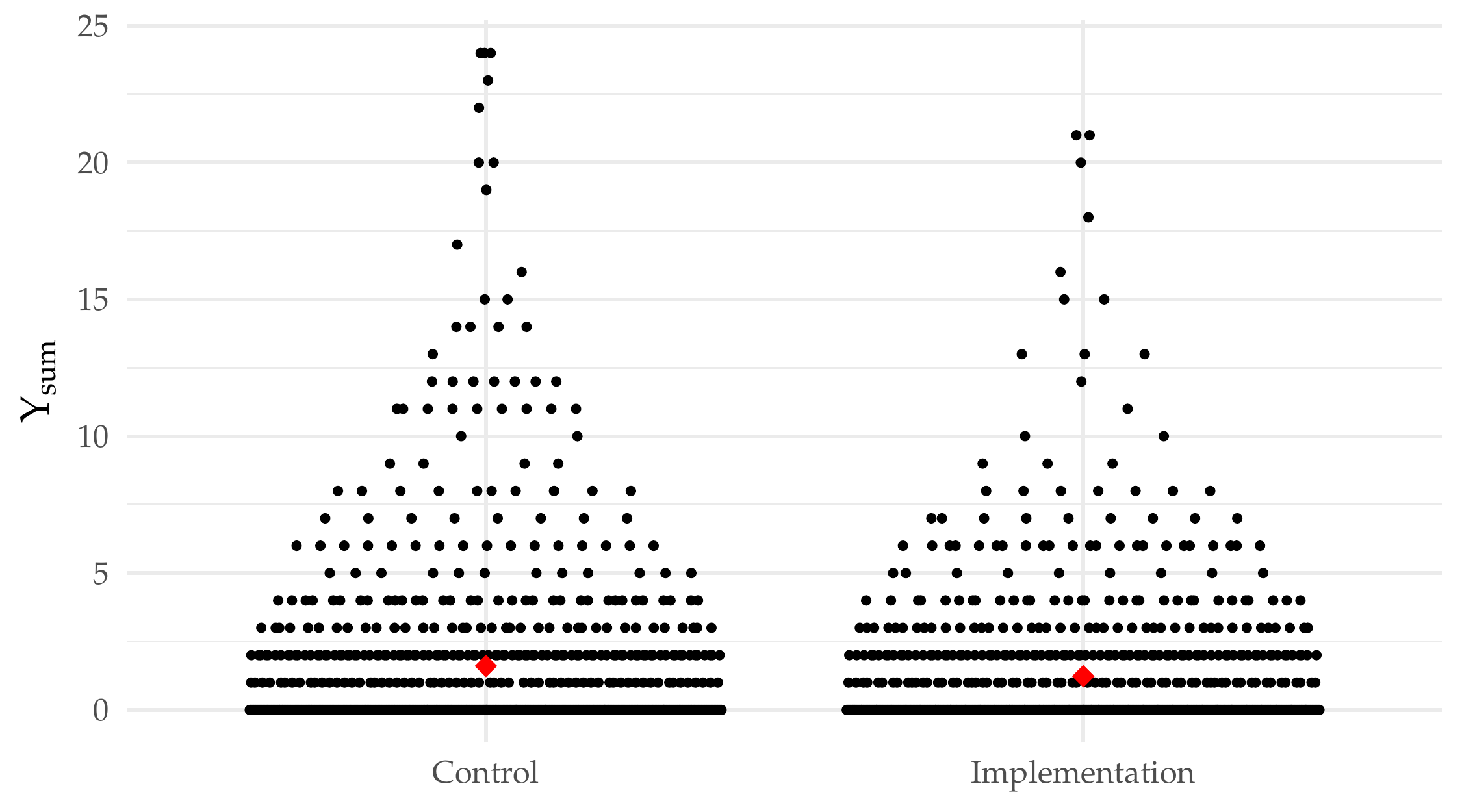}
 \caption{Differences in distribution of continuous sum violence measure at midline in the Becoming One trial. }
 \label{fig:b1_dist}
\end{figure}

\end{document}